\documentstyle[preprint,aps,epsfig]{revtex}

\begin{document}

\preprint{UTF/98-06-418; FTUV/98-64; IFIC/98-65}

\draft

\title{Polarized parton distributions and Light-Front Dynamics}
\author{{\large Pietro Faccioli}$^{a,}$\footnote{present address: 
ECT$\star$, Villa Tambosi, 38050 Villazzano,
(Trento), Italy}, {\large Marco Traini}$^{a,}$\footnote{
traini@science.unitn.it}, and
{\large Vicente Vento}$^{b,}$\footnote{vicente.vento@uv.es}}
\address{
$^a$ Dipartimento di Fisica, Universit\`a degli Studi di Trento,\\ 
and Istituto Nazionale di Fisica Nucleare, G.C. Trento\\
I-38050 POVO (Trento), Italy\\ 
$^b$ Departament de F\'{\i}sica Te\`orica, Universitat 
de Val\`encia\\
and Institut de F\'{\i}sica Corpuscular, Centre Mixt 
Universitat de Val\`encia,\\
Consejo Superior de Investigaciones Cient\'{\i}ficas,\\
E-46100 Burjassot (Val\`encia) Spain}
\maketitle

\begin{abstract}
We present a consistent calculation of the  structure functions
within a light-front constituent quark model of the nucleon. Relativistic
effects and the relevance of the covariance constraints are analyzed
for both polarized and unpolarized parton distributions. 
Various models, which differ in their gluonic structure at
the hadronic scale, are investigated. The results of the full
covariant calculation
are compared with those of a non-relativistic approximation to show 
the structure and magnitude of the differences.

 \end{abstract} 

\vskip 0.5cm
\leftline{Pacs: 12.39.Ki, 13.60.Hb, 14.20.Dh}
\leftline{Keywords: covariance, light-front, quark model, structure
functions, gluons}

\newpage

\section{Introduction} 
\label{intro} 
Constituent quark models, on one side, and the parton picture, on the
other side, represent two complementary descriptions of hadron
structure \cite{QandP}. The rest frame (or laboratory) description, based
on constituent quarks, and the infinite momentum frame description
($P_\infty$), based on partons, appeared as formalisms capable of
interpreting
particle physics phenomena at different resolution scales. The algebraic
approach, we are going to explore, seemed useful in establishing links
between them \cite{Melosh}.  It was intensively
investigated in the late sixties, but a quantitative scheme for
interacting scenarios was never attained.

The birth of the Quantum Chromodynamics (QCD) set the general framework to
understand deep inelastic phenomena beyond the parton model. The
perturbative approach to QCD is able to connect observables at different
resolution scales, but the realization of the complete project, i.e., to
fully understand the consequences of the dynamics of quarks and gluons,
requires the input of unknown non-perturbative matrix elements to provide
absolute values for the observables at any scale.

In the recent past the work by Gl\"uck, Reya and coworkers \cite{GRetal}
has shown that, starting from a parton parametrization at a low
resolution scale $\mu_0^2$, the experimental deep inelastic structure
functions at high momentum transfer can be reproduced and even predicted
\cite{GRprediction}.  $\mu_0^2$ is evaluated by evolving back the second 
moment of the valence distribution to the point where it becomes dominant.
The procedure, suggested by Parisi and Petronzio \cite{PaPetr}, assumes
that there exist a scale, $\mu_0^2$, where the short range (perturbative)
part of the interaction is negligible, therefore the glue and sea are
suppressed, and the long range (confining) part of the interaction
produces a proton composed of three (valence)
quarks only. Jaffe and Ross \cite{jaffeross80} proposed thereafter, to
adscribe the quark model calculations of matrix elements to that, hadronic
scale, $\mu_0^2$. For larger $Q^2$ their Wilson coefficients will give
the evolution as dictated by perturbative QCD. In this way quark models, 
summarizing a great deal of hadronic properties,
may substitute for low-energy parametrizations.   

Following such a path, a partonic description can be generated from gluon
radiation even off a purely valence quark system, which can be used
to generate the non perturbative input 
occurring in the Operator Product Expansion (OPE) analysis of 
lepton-hadron scattering in QCD \cite{evolving}.  A systematic analysis
based on non-relativistic potential models, shows that the approach can be
consistently developed at Next-to-Leading Order both for polarized and
unpolarized structure functions \cite{noi1}, including non-perturbative
contributions from the nucleon cloud \cite{MairTr} or from the partonic
substructure of the constituent quarks\cite{SVT97ss}.  The formulation of
this naive approach has been motivated by simplicity and by the success of
the constituent models in reproducing many properties of the hadronic
spectrum in the 1 - 2 GeV region.  

It has been quite evident, since the
original formulation of many of these models, that relativistic effects to
the nucleon wave function, as well as, covariance requirements, are needed
even for a phenomenological description of the structure of hadrons.
The present paper is devoted to the study of a possible generalization of
our approach in that direction, and we will show that relativistic
covariance can be incorporated within the same framework in a rather
transparent and elegant manner. To this aim we develop a constituent quark
model in the light-front realization of the Hamiltonian dynamics and apply
to it the formalism for the calculation of both polarized and unpolarized
parton distributions. 

In the next section we proceed to reformulate the procedure
using  the Hamiltonian formalism in the light front dynamics. A
light-front constituent quark model is defined and its momentum
distributions are related to the quark distribution function at the
hadronic scale in the conventional way. The third section is dedicated to
analyze, under different hadronic conditions, the perturbative evolution
required to reach the scale of the data. Two scenarios, characterized by
the presence or absence of soft gluons at the hadronic scale, are studied.
We end by summarizing our results and discussing those points, which at
the light of our presentation, require further development. 

\section{Quarks and Partons} 
\label{QPartons} 
A schematic phenomenological
distinction between constituent quarks and partons is often made by
differentiating the infinite momentum frame, where partons show up, from
the laboratory rest frame, where the constituent quarks appear to be the
relevant degrees of freedom responsible for the correct symmetries and
quantum numbers. Of course such a schematic point of view is only
approximate and can also be at the origin of some misunderstanding.
Actually the two descriptions appear well defined in the light-front
description of deep inelastic scattering, where the parton model is
recovered, in the Bjorken limit, due to the dominance of short light-cone
distances ($z^2 \sim 0$) in the relevant Feynmann diagrams. As a
consequence the partonic description can be developed in the rest frame of
the hadron by using light-cone formalism.  In particular the $i$-th parton
distribution can be related to the light-cone momentum density\footnote{A
formal derivation of Eq.~(\ref{DVal})  can be found in ref.\cite{noi1}.
Eq.~(\ref{DVal}) includes the support correction as discussed in the same
reference.}.  
\begin{equation} 
q_{\rm i}^{\uparrow\,(\downarrow)}(x,\mu_0^2) = {1 \over (1-x)^2}\, \int
d^3k\,\,n_{\rm i}^{\uparrow\,(\downarrow)}({\bf k}^2)\,\delta \left({x
\over 1-x} - {k^+\over M_N}\right)\,\,, \label{DVal} 
\end{equation} 
where $k^+/P^+=k^+/M_N=(\sqrt{{\bf k}^2+m_i^2}+k_z)/M_N$ is the light-cone
momentum fraction of the struck parton in the rest frame, $M_N$ and $m_i$
are the nucleon and parton mass respectively and $n_{\rm i}^\uparrow({\bf
k}^2)$, $n_{\rm i}^\downarrow({\bf k}^2)$ represent the light-cone density
momentum density of the $i$-th parton whose spin is {\it aligned}
($\uparrow$) or {\it anti-aligned} ($\downarrow$) to the total spin of the
parent nucleon. If one assumes that at the scale $\mu_0^2$ only the $u$
and $d$ constituent quarks are resolved, the momentum densities can be
written 
\begin{equation} 
n^{\uparrow\,(\downarrow)}_{u(d)}({\bf k}^2)=\langle
N,J_z=+1/2|\sum_{i=1}^3\, {1+(-)\tau_i^z \over 2}\,{1+(-)\sigma_i^z \over
2}\, \delta({\bf k}-{\bf k}_i)|N,J_z=+1/2\rangle\,\, .  \label{n_up/down}
\end{equation} 
The light-cone distributions (\ref{n_up/down}) have been
evaluated in the past making use of non-relativistic constituent quark
models, while in the present investigation we want to improve their
description including relativistic effects as introduced by a light-front
formulation of a three-body interacting system. As a consequence we will
remain within a constituent picture where the partons in the rest frame
are identified with three (constituent) quarks at the hadronic scale, and
covariance requirement is fulfilled.

Since the hadronic scale $\mu_0^2$ turns out to be very low  ($\mu_0^2
\sim (0.1 - 0.2)$ GeV$^2$), close to the constituent quark
mass\footnote{The
actual value of the scale $\mu_0^2$ ranges from 0.094 GeV$^2$, if only
valence quarks are considered, to 0.37 GeV$^2 \cite{MairTr}$,
when non-perturbative $q-\bar q$ pairs and gluons are included.}, we
assume that the constituent picture at this scale has to be
identified with a Constituent Quark Model, whose parameters are fixed
to reproduce the basic features of the nucleon spectrum in the
energy region $1 - 2$ GeV.  

\subsection{The Light-Front Constituent Quark Model} 
\label{LFCQM}

Before discussing the details of the quark wave functions, it is worth to
remark the main features of the theoretical framework that we have
employed to achieve a covariant description of the nucleon's constituent
degrees of freedom.  An exhaustive review of covariant hamiltonian
dynamics can be found in \cite{Keister}. 

On a quantum level, the covariance requirement is formulated by demanding
the probability assigned to a given event to be the same in all inertial
frames.  According to Wigner theorem \cite{Wigner1}, this is achieved if
(and only if)  the correspondence between the states in different frames
is realized by means of an unitary representation $U(\Lambda,a)$ of the
Poincar\'e group. 

The connected component of the Poincar\'e group is a ten parameter Lie
Group that includes four space time translations, three rotations and
three Lorentz boosts.  Let us denote with $P^{\mu}$ the representations of
the generators of space-time translations, with $K^j$ those of Lorentz
boosts, and with $ J^j$ those of $SU(2)$.  Any linear combination of
generators is still a generator.  To achieve a covariant formulation of
the Poincar\'e Algebra it is convenient to introduce a new set of tensor
generators of the Lorentz group:  
\begin{eqnarray}
 J^{0j}=K^j\\
 J^{jk}=\epsilon^{jkl}J_l\\
 J^{\alpha \beta}=-J^{\beta\alpha}.
\end{eqnarray}
The tensor character of these operators is expressed by the following
relationships:
\begin{eqnarray}
\label{tensor1}
U^{\dagger}(\Lambda)P^{\mu} U(\Lambda)= \Lambda^{\mu}_{\nu}P^{\nu}\\
U^{\dagger}(\Lambda)J^{\mu\nu}U(\Lambda) = \Lambda^{\mu}_{\rho}
\Lambda^{\nu}_{\sigma}J^{\rho\sigma}.
\label{tensor2}
\end{eqnarray}

Starting from the equations (\ref{tensor1}) and (\ref{tensor2})
 it is possible to work out the following Algebra :
\begin{equation}
[P^{\mu},P^{\nu}]  =  0 ,
\end{equation}
\begin{equation}
[J^{\mu\rho},P^{\nu}]  = i(g^{\mu\nu}P^{\rho}-g^{\rho\nu}P^{\mu}),
\end{equation}
\begin{equation}
[J^{\mu\nu},J^{\rho\sigma}]  =  i(g^{\mu\sigma}J^{\nu\rho}-g^{\mu\rho}J^{\nu\sigma}
-g^{\nu\sigma}J^{\mu\rho}+g^{\nu\rho}J^{\mu\sigma}).
\end{equation}

Hence, in order to build a covariant quantum mechanical model, one has 
to construct a model Hilbert space and find a set of operators  
such that these commutation relations are  satisfied.
 
In general, such a representation is not unique and depends on the
interaction. However, it is possible to single out a subset of generators
which do not contain the interaction. Such operators group together to
form the \emph{kinetic subgroup}. Once the kinetic subgroup is determined,
the whole set of generators is also fixed including the {\it Hamiltonian}
generators (i.e. those operators which contain the interaction).  Dirac
proposed \cite{Dirac} to fix first the kinetic subgroup identified with
the stability group of a three-dimensional surface in Minkowski's space,
and then extract the Hamiltonian generators by group representation
properties. 
 
The choice of the surface that determines the kinetic subgroup
fixes the {\it form of the dynamics}. In the following we will use the 
light-front form of the dynamics  whose invariant surface is the light-front
(null) plane $x^0+x^3=0$. This form of the dynamics has many important
properties, for example, it leads to the largest possible kinetic
subgroup.

In the light-front form of the dynamics, it is convenient to consider
 the following set of generators:  
$P^1$, $P^2$, $P^+=P^0+P^3$, $J^3$, $K^3$ (kinetic subgroup) 
and $P^-=P^0-P^3$,
${\bf F}_\perp= {\bf K}_\perp +\hat{{\bf z}}\times {\bf J}_\perp$
( set of Hamiltonians).
In particular, the role of the Hamiltonian $H$ of non relativistic
Quantum Mechanics is played, in the front form, by the
generator $P^-$ of translations along the direction orthogonal to
the null plane. This operator has a positive eigenvalued spectrum and this 
prevents the existence of negative "energy" states, 
allowing for the number of particles to be fixed. For this reason the
light-front form of the dynamics is particularly well suited to describe
a low energy model where only three valence quarks are considered. 

The construction of a representation of the Poincar\'e space such
that the kinetic subgroup is the stability group on 
the light-front, is clearly much 
easier by using a parametrization of the Minkowski 
space which is as close as possible 
to the symmetry of the problem, i.e. the light-front 
coordinates: $x^+=x^0+x^3$,
$x^-=x^0-x^3$, $x^1$ and $x^2$. In this way a vector 
is given by $(x^+, x^-, x^1, x^2) =
(x^+, x^-, {\bf x}_\perp)$, an the scalar product 
by $x\cdot y = -{1\over2} x^+ y^-
-{1\over2} x^- y^+ + {\bf x}_\perp \cdot {\bf y}_\perp$. 
The null plane is the plane 
with $x^+ = 0$ and the "(three)-vector" $(x^-, {\bf x}_\perp)$ 
on that plane is a true
three-vector under light-front boosts (i.e. the boost belonging 
to the kinetic subgroup).

In order to model the Hilbert space one needs to combine the generators of 
the Poincar\'e group to get a set of self commuting
operators which have a physical interpretation. 
By combining $P^\mu$, $J^{\mu\nu}$ one can construct two invariants:
\newline
i) the mass operator $$
M^2 = -P^2 = P^+ P^- -{\bf P}_\perp^2\,,
$$
ii) and the square ($W_\mu\,W^\mu$) of the {\it Pauli-Lubansky} vector:
$$
W_{\mu} = {1\over2} P^\nu J^{\rho\sigma}\epsilon_{\nu\rho\sigma\mu}\,.
$$

In relativistic systems the definition of the  
spin operator is related to the 
Pauli-Lubansky vector: one defines, in fact, the square of the spin by
\begin{equation}
\label{lubaspin}
W^2=-M^2j^2.
\end{equation}
It can be proved that there is an infinite set of operator-valued vectors
satisfying the relation (\ref{lubaspin}).
Therefore, the spin operator is defined 
as a linear combination of $W_{\mu}$ and $M^{-1}$:
\begin{equation}
j_\alpha = u_\alpha^\mu(P) W_\mu\,M^{-1}\,.
\label{spindef}
\end{equation}
The coefficients $u_\alpha^\mu(P)$ are space-like 
operator-valued vectors of an orthonormal basis  \cite{lf}.
Different choices of this basis leads to the specific definition of spin
 in different forms of the dynamics.
Starting from (\ref{spindef}) one can prove 
that the spin operator transforms as
\begin{equation}
U^\dag(\Lambda)\,\vec j\,U(\Lambda) = R_W(\Lambda,P)\,\vec j\,,
\label{spintransf}
\end{equation}
where $R_W(\Lambda,P)_{\alpha \beta} = U_\alpha(\Lambda P) \cdot \Lambda\,u_\beta(P)$
is called Wigner rotation.
One can demonstrate that defining
$u_0(P) = P/M$, the 16 components quantity $u_\mu^\nu(P)$ forms a $SO(1,3)$ 
representation
of a Lorentz transformation. The spin operator can be 
written
\begin{equation}
j_\alpha = {1\over 2}\,\epsilon_{0\alpha\beta\gamma}\,
u^\beta_\rho(P)\,u_\sigma^\gamma(P)\,J^{\rho\sigma} = 
\frac{1}{2} \epsilon_{0\alpha\beta\gamma} u^\beta_\rho(P)\,u_\sigma^\gamma(P)\,
\sum_i^n J_i^{\rho\sigma}\,,
\label{spindef1}
\end{equation}
where the last equality holds when the system is made up of $n$ independent 
constituents. In the {\it instant form} of the dynamics (in which the
invariant surface is chosen to be the equal-time plane)
the Wigner transformations associated to a rotation is the rotation itself and 
the spin is called {\it canonical}. One can demonstrate that, 
in this form of the 
dynamics, the ordinary composition rules of the angular momentum 
hold, providing one rotates the constituent spins 
by means of appropriate Wigner 
rotations. In the case of light-front form of the 
dynamics, the Wigner rotations 
associated to a light-front boosts are the identity. 
The ordinary composition laws 
for the total angular momentum hold, provided we 
transform the light-front spin
into the canonical spin. The transformation that 
relates the light front spin to the
canonical spin is called  a Melosh rotation,
\begin{equation}
R_M(P)_{\alpha\beta}:=u^{\mu}_\alpha (P)^{can.} 
u^{\nu}_{\beta}(P)^{l.f.} g_{\mu\nu}
\end{equation}
Its spin-$\frac{1}{2}$ representation  is:
\begin{equation}
D^{1\over2}\left[R_M({\bf P},M)\right]=
{M+P^+-i\vec \sigma \cdot (\hat z \times {\bf P}_\perp)
\over \sqrt{(M+P^+)^2+{\bf P}_\perp^2}}\,\,.
\label{melosh1}
\end{equation}

\subsubsection{Non-interacting and interacting systems on the light-front}

The single spin-$1/2$ particle states can be described by wave functions 
$\phi(\tilde p, \lambda)$, where 
$\tilde p = (p^+,{\bf p}_\perp)$ is the momentum 
three-vector on the light-front and $\lambda$ the longitudinal spin component.
One has:
\begin{eqnarray}
||\phi||^2 & = & \sum_{\lambda = -j}^{\lambda = +j}\,\int 
{d^3[\tilde p]\over p^+}\,
|\phi(p^+,{\bf p}_\perp)|^2\,, \nonumber \\
d^3[\tilde p] & = & d p^+\,d^2 {\bf p}_\perp\,\theta(p^+)\,;
\label{measure}
\end{eqnarray}
where the measure $d^3[\tilde p]/ p^+$ and the 
component $\lambda$ are invariant 
under the kinematic transformations, 
and the factor $1/\sqrt{p^+}$ can be included in the 
definition of the wave function provided we change its light-front boost 
transformation properties  accordingly.

For a many-body system of $n$ non-interacting particles the wave function 
can be written $\Phi(\tilde p_1,\lambda_1;...;\tilde p_n,\lambda_n)$, and 
Eq.~(\ref{measure}) becomes
\begin{equation}
||\Phi||^2 = \sum_{\lambda_1,...,\lambda_n}\,\int\,
\left(\prod_{i=1}^n {d^3[\tilde p_i]\over p_i^+}\right)\,
|\Phi(\tilde p_1,\lambda_1;...;\tilde p_n,\lambda_n)|^2\,\,.
\label{measure_n}
\end{equation}

Quite often it is more useful to represent $\Phi$ as 
functions of the total momentum 
on the light-front $\tilde P=\sum_{i=1}^n \tilde p_i$, the momentum fraction
$x_i=p_i^+ / P^+$,  and the transverse relative momenta 
${\bf k}_{i \perp}=u_\perp \cdot p_i = {\bf p}_{i \perp} - x_i {\bf P}_\perp$.
Eq.~(\ref{measure_n}) can be then transformed into
\begin{eqnarray}
||\Phi||^2 & = & \sum_{\lambda_1,...,\lambda_n}\,\int\,
{d^3[\tilde P]\over P^+}\,
\int\,\left(\prod_{i=1}^n {d x_i\over x_i}\, d^2 {\bf k}_{i \perp}\right)\,
\delta \left(\sum_{i=1}^n x_i -1 \right) \delta \left(\sum_{i=1}^n  
{\bf k}_{i \perp} \right)
\times \nonumber \\
& \times & |\Phi(\tilde P, {\bf k}_{1 \perp},x_1,\lambda_1;...; 
{\bf k}_{n \perp},x_n,\lambda_n)|^2\,\,.
\label{transv_n}
\end{eqnarray}
One can introduce a further set of coordinates adding, 
on kinematical basis only, 
a third component $k_{i 3}$ to the transverse 
momenta ${\bf k}_{i \perp}$ boosting 
the lab momenta $p_i$:
\begin{eqnarray}
k_{i 3} = u_3(P) \cdot p_i & = & {1\over2}\,
\left[M_0 x_i + {m_i^2 + {\bf k}^2_{i \perp}
\over M_0 x_i}\right] = M_0 x_i - \omega_i\,\,,\nonumber \\
\omega_i = -u_0(P) \cdot p_i & = & {1\over2}\,
\left[M_0 x_i - {m_i^2 + {\bf k}^2_{i \perp}
\over M_0 x_i}\right]\,\,,
\end{eqnarray}
where the total mass operator of the system
\begin{equation}
M_0 = \sum_i \sqrt{{\bf p}_i^2 + m_i^2} = 
\sqrt{ \sum_i {{\bf k}^2_{i \perp} + m_i^2 \over x_i}} =
\sum_i \omega_i = \sum_i \sqrt{{\bf k}_i^2 + m_i^2}\,\,.
\label{freemass}
\end{equation}
The invariant measure to be used in the definition of the scalar product
transforms in
the following way:
\begin{equation}
\left(\prod_{i=1}^n {d x_i\over x_i}\, d^2 {\bf k}_{i \perp}\,\right)
\delta \left(\sum_{i=1}^n x_i -1 \right) \delta 
\left(\sum_{i=1}^n  {\bf k}_{i \perp} \right)
\to  \left(\prod_{i=1}^n {d^3 {\bf k}_i\over \omega_i}\right)\,M_0\,
\delta \left(\sum_{i=1}^n  {\bf k}_{i} \right)\,\,.
\end{equation}

The total spin, in the new coordinate system, is
\begin{equation}
\vec j = \sum_{l=1}^n \left[i\,\vec \nabla_{{\bf k}_l} \times {\bf k}_l + \vec s_l \right]\,\,,
\end{equation}
with
\begin{equation}
\vec s_l = R_M({\bf k}_l,m_l)\,\vec j_l 
\label{spinp}
\end{equation}
and the representation of the Melosh rotation reads
\begin{equation}
D^{1\over2}\left[R_M({\bf k}_i,m_i)\right]={m_i + x_i M_0-i\vec \sigma 
\cdot (\hat z 
\times {\bf k}_{i \perp})
\over \sqrt{(m_i+x_i M_0)^2+{\bf k}_{i \perp}^2}}\,\,.
\label{melosh2}
\end{equation}
As a first consequence of this section we can conclude that the dynamics of $n$ 
non-interacting spin $1/2$ particles can be described, on the light-front, by 
a wave function  $\psi(\tilde P,{\bf k}_1,\mu_1;...;{\bf k}_n,\mu_n)$
whose norm is given by
\begin{equation}
||\psi||^2 = \langle \psi|\psi\rangle =
\sum_{\mu_1,...,\mu_n}\,\int\,{d^3 \tilde P \over P^+}\,
\left(\prod_{i=1}^n {d^3 \bf k}_i\right)\,\delta \left(\sum_{i=1}^n  
{\bf k}_{i} \right)
\,|\psi|^2\,\,,
\label{normpsi}
\end{equation}
where $\mu_i$ are the eigenvalues of the spin projections $s_i^z$ of
eq.(\ref{spinp}).
We stress that the Jacobian and the invariant measure of the transformation
$({\bf k}_{i \perp},x_i) \to {\bf k}_i$, has been absorbed in the definition 
of the wave functions $\psi$ as can be seen from Eq.~(\ref{normpsi})
(cfr. the discussion after Eq.~(\ref{measure})).

\vspace{3mm}

The extension to interacting systems requires a dynamical representation of
the Poincar\`e group.
One way to achieve this is to  add an interaction
$V$ to the free mass operator
$M_0$
\begin{equation}
\label{eigmasseq}
M = M_0 + V\,\,,
\label{massV}
\end{equation}
where $V$ is the interaction. All the other 
definitions remain unaffected, including 
the definition of the third component of 
the relative momenta $k_{i 3}$, because the 
boosts we use are interaction independent. 
All required commutation relations are satisfied 
if the mass operator commutes
with the total spin $\vec{j}$ and with the kinematic generators.
In the representation in which states are represented by functions of
$\tilde{P}$, ${\bf k}_i$ and $\mu_i$ these conditions are realized if: 

i) $V$ is independent on the total momentum ${\bf \tilde{P}}$; 

ii)$V$ is invariant under ordinary rotations.

\vspace{3mm}

{\it Summarizing}: in the light-front  formulation of the quark
dynamics, the intrinsic momenta of the constituent quarks ($k_i$) can be 
obtained from the corresponding momenta ($p_i$) in a generic reference 
frame through a light-front boost 
($k_i = u(P)\cdot p_i$, $P \equiv \sum_{i=1}^3\,p_i$) 
such that the Wigner rotations reduce to the identity.
The spin and spatial degrees of freedom are described by the wave
function:
\begin{equation}
\Psi=\frac{1}{\sqrt{P^+}}\,\delta(\tilde{P}-\tilde{p})\,
\chi({\bf k}_1,\mu_1,\dots,{\bf k}_3,\mu_3),
\end{equation}
where $\mu_i$ refer to the eigenvalue of the light-front spin, 
 so that the spin part
of the wave function is transformed by the tensor product of three
independent Melosh rotations: $R^{\dag}_{\rm M}({\bf k}_i, m_i)$
\cite{Melosh}, namely ${\cal R}^{\dag} = \prod_{\otimes,
i=1}^3\,R^{\dag}_{\rm M}({\bf k}_i, m_i)$.

The internal wave function is an eigenstate of the baryon mass operator
$M=M_0+V$, with  $M_0 = \sum_{i=1}^3\,\omega_i =
\sum_{i=1}^3\,\sqrt{{\bf k}_i^2+m_i^2}$,
where the interaction term $V$ must be independent on the total momentum
$\tilde{P}$ and invariant under rotations.

The nucleon state is then characterized by isospin (and its third
component), parity, light-front (non-interacting) angular momentum operators 
$J$ and projection $J_n$, where the unitary vector 
$\hat n = (0,0,1)$ defines the spin
quantization axis.

\subsection{Valence Quark Hamiltonian}
\label{VQHam}

In the present work we will discuss results of a confining mass equation 
(\ref{eigmasseq}) of the following kind
\begin{equation}
\left(M_0 + V\right)\,\psi_{0,0}(\xi) \equiv \left(\sum_{i=1}^3\,
\sqrt{{\bf k}_i^2+m_i^2} -{\tau \over \xi} + \kappa_l \,\xi\,\right)
\,\psi_{0,0}(\xi) = M\,\psi_{0,0}(\xi)\,\,,
\label{massop}
\end{equation}
where $\sum_i {\bf k}_i = 0$, $\xi = \sqrt{\vec \rho\,^2 + \vec \lambda\,^2}$ 
is the radius of the hyper-sphere in six dimension and $\vec \rho$ and 
$\vec \lambda$ are the intrinsic Jacobi coordinates 
$\vec \rho = ({\bf r}_1 - {\bf r}_2)/\sqrt2$, 
$\vec \lambda =({\bf r}_1 + {\bf r}_2 -2\,{\bf r}_3)/\sqrt6$. 
The choice of the mass operator (\ref{massop}) has a twofold motivation:

\noindent i) Simplicity. This is a first attempt to develop a covariant 
approach to DIS based on quark models, therefore we use a mass operator
whose symmetry
properties facilitate the numerical solutions. In this respect the choice 
of a hypercentral potential has big advantages.

\noindent ii) The nucleon spectrum is reproduced. 
This has been  demonstrated both within 
non-relativistic \cite{TBM95} and relativistic \cite{CoeDaRi97} approaches
\footnote{In ref.\cite{CoeDaRi97} a simplified version of the 
same mass operator has been introduced to investigate the baryonic 
mass spectrum in the instant form dynamics. Namely the eigenstates of the
operator $\hat M^2 = \sum_i ({\bf k}_i^2 + m_i^2) -a/\xi + b\,\xi$ are 
discussed, and the potential $-a/\xi + b\,\xi$ is the simplified form of 
the more complex expression $V^2 + \{M_0,V\} +
\sum_{i\neq j} \sqrt{{\bf k}_i^2+m_i^2}\,\sqrt{{\bf k}_j^2+m_j^2}$
obtained by squaring the correct mass operator $\hat M = M_0+V$. We will 
discuss solutions of the mass operator 
Eq.~(\ref{massop}) directly, without further simplification.}.
In particular the well known problem of the mass of the Roper is 
solved in the present case by the use of $\tau/\xi$ potential, 
as discussed by Ferraris {\it et al.} \cite{TBM95}. Of course our aim 
is much less ambitious than reproducing
all the complexity of the baryon spectrum, we restrict the calculation to 
the nucleon wave function and we do not consider spin-dependent terms 
assuming pure $SU(6)$-symmetric states\footnote{Adding a 
perturbative hyperfine interaction as discussed in ref.\cite{TBM95} 
in the context 
of non-relativistic Hamiltonians, would be 
rather simple also within the relativized
scheme of Eq.~(\ref{massop}). 
At the present stage we neglect such a complication, 
but a more complete study of the nucleon 
spectrum should include $SU(6)$-breaking terms 
within the light-front approach.}.

The intrinsic nucleon state is antisymmetric in the color degrees of
freedom and symmetric with respect the orbital, spin and flavor coordinates. 
In particular, disregarding the color part, one can write
\begin{equation}
|N, J, J_n = +1/2 \rangle = 
\psi_{0,0}(\xi)\,{\cal Y}\,^{(0,0)}_{[0,0,0]}(\Omega)\,
\,{\chi_{MS} \phi_{MS} + \chi_{MA} \phi_{MA}\over \sqrt {2}}\,\,,
\end{equation}
where $\psi_{\gamma,\nu}(\xi)$ is the hyper-radial wave function solution of
Eq.~(\ref{massop}), ${\cal Y}\,^{(L,M)}_{[\gamma,l_\rho,l_\lambda]}(\Omega)$
the hyper-spherical harmonics defined in the hyper-sphere of unitary radius,
and $\phi$ and $\chi$ the flavor and spin wave function of mixed $SU(2)$
symmetry. Let us note that, in order to preserve relativistic covariance, 
the spin wave functions
\begin{equation}
\chi_{MS} = {1 \over \sqrt 6}\,
\left[2\,\uparrow \uparrow \downarrow - \left(\uparrow \downarrow + 
\downarrow \uparrow\right)\uparrow\right]\;;
\hspace{10mm}
\chi_{MA} = {1\over \sqrt 2}\,
\left(\uparrow \downarrow - \downarrow \uparrow\right)\uparrow
\label{spinwf} \,\,,
\end{equation}
have to be formulated by means of the 
appropriate Melosh transformation of the
i-$th$ quark spin wave function:
\begin{equation}
\uparrow_i \; \equiv \; R_{\rm M}({\bf k}_i, m_i)  
\left(
\begin{array}{c}
1 \\
0 
\end{array}
\right)
\; = \; 
{1 \over \sqrt{(m_i+x_i M_0)^2 + {\bf k}_{i\,\perp}^2}}\;
\left(
\begin{array}{c}
m_i + x_i\,M_0 \\
k_{i\,R}
\end{array}
\right)
\label{meloshs1}
\end{equation}
\begin{equation}
\downarrow_i \; \equiv \; R_{\rm M}({\bf k}_i, m_i)  
\left(
\begin{array}{c}
0 \\
1 
\end{array}
\right)
\; = \; 
{1 \over \sqrt{(m_i+x_i M_0)^2 + {\bf k}_{i\,\perp}^2}}\;
\left(
\begin{array}{c}
- k_{i\,L} \\
m_i + x_i\,M_0 
\end{array}
\right)
\label{meloshs2}
\end{equation}
where $x_i = p^+_i/P^+ = (k_{i\,z} + \omega_i)/M_0$ and
$k_{L/R} = k_x \pm\,i\, k_y$. 

\subsubsection{Numerical Solutions}
\label{numsol}

We have solved the mass equation (\ref{massop}) numerically by expanding
the hyper-radial wave functions $\psi_{\gamma \nu}(\xi)$ on a truncated set 
of hyper-harmonic oscillator basis states. The matrix elements of the mass
operator (\ref{massop}) have been calculated in momentum space 
for the kinetic energy term, and in configuration space for the
interaction 
one.  Making use of  the Rayleigh-Ritz variational principle the HO
constant
has been determined and convergence has been reached considering a basis as
large as 17 HO components \cite{TrFaVe}. The parameters of the
interaction, have been determined phenomenologically in order to reproduce 
the basic features of the (non strange) baryonic spectrum up to $\approx 1500$
MeV, namely the position of the Roper resonance and the average value of the $1^-$
states.
We obtain: $\tau = 3.3$ and $\kappa_l = 1.8$ fm$^{-2}$\cite{TrFaVe} to be 
compared with the corresponding non-relativistic fit $\tau = 4.59$ and 
$\kappa_l = 1.61$ fm$^{-2}$\cite{TBM95}. The constituent quark masses have been
chosen $m_u = m_d = m_q = M_N/3$. 

The result of this reparametrization is that  a huge amount of high
momentum components is generated by solving the mass equation (cfr.Fig.~1a),
and they play an important role in the evaluation of transition and 
elastic form  factors within light-front constituent quark models as
discussed by Caldarelli {\it et al.}\cite{romalf} in connection with 
the solutions of the Isgur-Capstick model Hamiltonian.
From the point of view of DIS the enhancement of high
momentum components in the density distribution of the valence quarks is
crucial to reproduce the behavior of the unpolarized structure functions 
for large value of the Bjorken variable $x$ as it will be discussed in 
the next section where also polarized responses are investigated at the
hadronic scale.

\subsection{Partons at the Hadronic Scale}

The calculation of the partonic content at the hadronic scale 
as given by Eq.~(\ref{DVal}), is rather involved mainly because of 
the spin dynamics accounted for by the Melosh rotations (\ref{meloshs1}) 
and (\ref{meloshs2}), but finally the polarized distributions at the 
hadronic scale  can be written in a rather elegant way:
\begin{equation}
{\cal Q}(x,\mu_0^2) = 
{\pi\over 9}\,{M_N \over (1-x)^2}\,
\int_0^\infty dt\,n(\tilde k_z^2,t)\,{a_{\cal Q}\,\left(
m + \sqrt{\tilde k^2_z + t + m^2}+ \tilde k_z\right)^2 
+ b_{\cal Q}\,t \over
\left(m + \sqrt{\tilde k^2_z+t+m^2}\right)^2 + t}\,D(\tilde k_z,t)
\label{q-melosh}
\end{equation}
where ${\cal Q} \equiv u^{\uparrow\,(\downarrow)}, d^{\uparrow\,(\downarrow)}$, 
$t \equiv {\bf k}^2_{\perp}$,
\begin{equation}
D(\tilde k_z,t) = {{\sqrt{t+\tilde k_z^2+m^2}\over \left|
\sqrt{t+\tilde k_z^2+m^2} + \tilde k_z\right|}}\,\,,
\end{equation}
$$
\tilde k_z(x,t) = {M_N\over 2}\,\left[{x \over 1-x}-{(t+m^2)\over M_N^2}\,
{(1-x)\over x}\right]\,\,,
$$ 
and  $n(\tilde k_z^2, {\bf k}^2_{\perp})$ is the total
momentum density distribution of the valence quarks in the nucleon
calculated making use of the eigenfunction $\psi_{0\,0}$ of 
Eq.~(\ref{massop}):
\begin{equation}
n({\bf k}^2) = 3\,\int|\psi_{0\,0}|^2\,\delta({\bf k}-{\bf
k}_3)\,\delta(\sum_i {\bf k}_i)\,d^3{\bf k}_1 d^3{\bf k}_2 d^3{\bf k}_3\,\,,
\label{n_of_k}
\end{equation}
with $\int d^3{\bf k}\, n({\bf k}^2) = 3$.
The coefficients $a_{\cal Q}$ and $b_{\cal Q}$ take the values
$a_{u^{\uparrow}} = b_{u^{\downarrow}} = 5$;
$b_{d^{\uparrow}} = a_{d^{\downarrow}} = 2$; 
$b_{u^{\uparrow}} = a_{u^{\downarrow}} = a_{d^{\uparrow}} = b_{d^{\downarrow}} = 1$.

\subsubsection{Effects of Melosh Rotations}

The effects of high momentum components on the unpolarized parton distributions 
$\sum_{q=(u,d)}\,(q^\uparrow(x,\mu_0^2)+q^\downarrow(x,\mu_0^2)) = 
u_V(x,\mu_0^2) + d_V(x,\mu_0^2)$
at the hadronic scale are shown in Fig.~1b. Their important r\^ole in 
reproducing the correct behavior of the structure functions for large
values of the Bjorken variable $x$ will be discussed in the next sections.
The relevant effects of relativistic covariance are even 
more evident looking at the polarized distributions
$\Delta u_V(x,\mu_0^2)= u^\uparrow(x,\mu_0^2)-u^\downarrow(x,\mu_0^2))$,
$\Delta d_V(x,\mu_0^2)= d^\uparrow(x,\mu_0^2)-d^\downarrow(x,\mu_0^2))$ 
where the spin dynamics on the light-front plays a crucial r\^ole. The 
introduction of the Melosh rotations results in a substantial enhancement 
of the responses at large $x$ and in an suppresion of the response for
$0.1 \lesssim x \lesssim 0.5$ as can be seen from Fig.~2. We show, in the
same figure, 
also the predictions of a pure relativized solution obtained by solving 
numerically Eq.~(\ref{massop}) and neglecting the Melosh rotation effects 
in (\ref{q-melosh}). Such a calculation retains the contribution due to
the high momentum components, while the covariance requirement on the 
parton distribution is lost. One can argue on the quite relevant r\^ole of
covariance on the spin observables.

\subsubsection{Non-perturbative Gluons}
\label{nonpgluo}

A natural choice for the unpolarized gluon distribution within 
the present approach, has been discussed in refs.\cite{noi1} and 
assumes a {\it valence-like} form 
(cfr. also ref.\cite{GRetal})
\begin{equation}
G(x,\mu_0^2)= {{\cal N}_g \over 3}\,\left[ u_V(x,\mu_0^2) +
d_V(x,\mu_0^2)\right]\,\,.
\label{nonpg}
\end{equation}
This definition implies $\int G(x,\mu_0^2)\,dx = 2$ and therefore only 
$60\%$ of the total momentum is carried by the valence quarks at the 
scale $\mu_0^2$. In this case $\mu_0^2 = 0.220$ GeV$^2$ at NLO 
([$\alpha_s(\mu_0^2)/(4\,\pi)]_{\rm NLO} = 0.053$).

If the gluons were fully polarized at the scale $\mu_0^2$ one would
have  $|\Delta G(x,\mu_0^2)| = G(x,\mu_0^2)$. 
However recent next-to-leading (NLO) studies of polarized Deep Inelastic
Scattering (DIS) 
\cite{BaFoRi96,workshop97,GS96,GRSV96,deFS97,LPS97,BFR&A96/97} 
show that the $x$-shape of $\Delta G$ is hardly constrained 
by the present data. As a consequence quite a few sets of parton 
distributions, with rather different $x$-shapes of the gluon component, 
provide very good descriptions of the same data. In particular the sign 
of $\Delta G(x,Q^2)$ for $x \gtrsim 0.1$ is not convincingly fixed: in the 
parametrization $C$ of ref.\cite{GS96} it is assumed to be {\it negative}, 
while it is taken {\it positive} in other cases (cfr., for example, 
ref.\cite{GRSV96}).

In the recent past different authors have discussed the {\it sign} of the 
gluon contribution to the nucleon's spin in the light-cone gauge 
($A^+ = 0$) and at the hadronic scale valid at the quark model level. 
In particular Ball, Forte and Ridolfi \cite{BaFoRi96} showed the 
compatibility of the experimental data with a large and 
{\it positive} gluon polarization at $Q^2 = 1$ GeV$^2$ 
($\Delta G(Q^2) = \int_0^1 dx \Delta G(x,Q^2) \approx 1.6\,\pm\,0.9$); 
Jaffe \cite{jaffe96} showed that the color electric and magnetic fields, which 
are believed to be responsible for the spin splittings among light baryons, 
give rise to a significant {\it negative} contribution 
($\Delta G(\mu_0^2) \approx - 0.7$) at the hadronic 
scale; Mankiewicz, Piller and Saalfeld \cite{MaPiSaa97} performed 
a $QCD$ sum rule calculation  obtaining $\Delta G(\mu_0^2) \approx 2.1\,\pm\,1.0$, 
while Barone, Calarco and Drago \cite{BaCaDra98}, by using the non-relativistic 
quark model of Isgur-Karl and considering the renormalization due to 
self-interaction contributions, obtained $\Delta G(\mu_0^2 = 0.25$ GeV$^2) 
\approx 0.24$.

We shall describe results in two scenarios characterized by different 
gluon distributions ($\Delta G$) at the hadronic scale \cite{noi1}:
\begin{itemize}
\item [i)] Scenario A : $\Delta G(x,\mu_0^2) = 0$.  
Only quark valence distributions are allowed at the hadronic scale. 
The momentum sum rule determines $\mu_0^2 = 0.094$ GeV$^2$
at NLO ([$\alpha_s(\mu_0^2)/(4\,\pi)]_{\rm NLO} = 0.142$).

\item [ii)] Scenario B: $\Delta G(x,\mu_0^2) = f\, G(x,\mu_0^2)$.
$f$ is the fraction of polarized gluons and has to be considered with
the appropriate sign. For example the Jaffe's suggestion, in our 
approximation scheme, would imply $\Delta G(x,\mu_0^2) 
\approx - 0.35\,G(x,\mu_0^2)$, while $f > 0$ will imply positively 
polarized gluons
\footnote{In fact in ref.\cite{jaffe96} it has
been shown that $\int \Delta G(x,\mu_0^2)\,dx < 0$. Such inequality 
does not imply $\Delta G(x,\mu_0^2) < 0$ in the whole $x$-range, and our results 
for positively and negatvely polarized gluons can be seen as upper and lower limits
for both cases.}.

\end{itemize}

\section{Results}
\label{res}

\subsection{Partonic bremsstrahlung}

An important aspect of our approach is  that we perform, 
in both the flavor non-singlet and singlet channels, a complete analysis at 
Next-to-Leading Order, an essential requirement needed to include the 
anomalous $(\gamma_5)$ contributions consistently. Complete NLO calculation of all
polarized two-loop splitting functions \cite{polNLO96}, 
in the $\overline {MS}$ renormalization 
and various factorization schemes\cite{LPS97,BFR&A96/97,polNLO96}), are now 
available. The evolution of the polarized distributions 
is performed according to the 
solution of the renormalization group equation inverting the moments
$\langle f(Q^2)\rangle_n = \int_0^1 dx\,f(x,Q^2)\,x^{n-1}$ numerically.
Since the starting point for the evolution ($\mu_0^2$), consistent with 
the renormalization scale of the quark model, is rather low, 
the form of the equations must guarantee complete symmetry for the evolution 
from $\mu_0^2$ to $Q^2 \gg \mu_0^2$ and {\it back~}
avoiding additional approximations associated with Taylor expansions and not
with the genuine perturbative $QCD$ expansion 
(a complete discussion on the evolution procedure can be found in 
ref.\cite{noi1}).

We use the so called $\overline {MS}$ and AB factorization scheme
evolutions on our model input to show that our results are largely
independent of which one we use. In principle the calculation should be
independent of factorization scheme, however at present it is not
straightforward to abscribe the non-perturbative input to any specific
factorization scheme. The spurious factorization dependence arises
because we associate the same non-perturbative input to the different
factorization schemes as discussed in the next sections.

\subsection{Discussion}

\subsubsection{Polarized Parton Distributions}

In Fig.~3 the results for the proton structure function $g_1^p(x,Q^2)$ are
shown and compared with the experimental data. The non-relativistic
approximation of the present calculation appears to reproduce rather poorly 
the experimental observations. Even the use of non-relativistic models
which reproduce rather well the whole baryon spectrum does not alter this
conclusion, as shown in ref.\cite{noi1} where the predictions of quite a few
non-relativistic models are discussed.

The full covariant calculation leads to a theoretical
predictions quite close to experimental data in the region 
$0.01 \leq x \lesssim 0.4$ even within the simple assumption 
of a pure valence component at the hadronic scale 
(scenario A). The calculation is parameter-free and the only 
adjustable parameters of the Hamiltonian ($\tau$ and $\kappa_l$ 
in Eq.~(\ref{massop})) have been fixed to reproduce the low-lying 
nucleon spectrum as discussed in section \ref{numsol}.

The effect of relativistic covariance in the quark wave function is mainly 
associated to the spin dynamics induced by the Melosh rotations 
(cfr. Eqs.(\ref{meloshs1}), (\ref{meloshs2}),
(\ref{q-melosh})) and these transformations lead to a
strong suppression of the structure function in
the small-$x$ region ($x \lesssim 0.5$).

In order to introduce the gluons non perturbatively we evolve the
unpolarized distributions predicted by the scenario A, up to the scale 
of scenario B where 60\% of the total 
momentum is carried by valence quarks. At that scale we add to 
the valence partons the non-perturbative gluons as defined in 
Eq.~(\ref{nonpg}) of section \ref{nonpgluo}.

Moreover, at the new hadronic scale, the fraction of polarized gluons 
is chosen to be {\it negative} according to Jaffe's result \cite{jaffe96}
or {\it positive} as in other parametrizations. 
Note that by choosing a negative fraction we are maximizing the difference
with respect to the radiative gluons,
because those yield to a positive polarization. Fig.~6 leads us to
confirm that the low-$x$ data on $g_1^p$ do not constrain the gluon
polarization strongly. If we vary the fraction of polarized gluons from 35\% to
100\% the quality of the agreement is not significantly changed in the region 
$0.01 \leq x \lesssim 0.4$. For larger values of $x$ the valence contribution 
plays a major role and the behavior of the structure functions might 
depend
on the model wave functions.

For both  scenarios the boundary value $\mu_0^2$ is rather low and one 
could question on higher order contributions to the evolution. In order to
appreciate (and maximize) the relevance of NNLO (or higher order) effects,
the polarized proton structure function are evolved at $Q^2=3$ GeV$^2$
within the scenario A and by using both LO and NLO 
approximations (cfr. Fig.~4). In the region of $0.01 \lesssim x \lesssim
0.5$ they differ roughly of 20\% which represents the order of magnitude 
of the largest higher leading effects. The same evolution obtained within
the scenario B would lead to an even smaller
difference ($\lesssim 10\%$) between LO and NLO results.

The comparison of the predicted neutron structure function with
the data (Fig.~5) differs quite substantially according to the amount of 
polarized gluons at the hadronic scale. Within scenario A the values of 
$g_1^n(x,Q^2)$ remain quite small according to the fact that the mass operator 
chosen (\ref{massop}) is $SU(6)$ symmetric. The introduction of {\it
negative} gluon polarization brings the predictions of the present
relativistic quark model closer to the experimental observations at least
in the $x \geq 0.1$ region; a larger {\it negative} fraction of gluon 
polarization is favored by the data as can be argued also from the results
of {\it positively} polarized gluons as shown in the same figure. 
Any how, since the neutron structure function is largely dependent 
on the nucleon wave function model, our assumption on $SU(6)$ symmetry 
is the largest source of uncertainty and a more detailed study of 
$g_1^n$ should include more complex wave function models.  Another 
source of uncertainty in our evaluation of the neutron stucture function 
is the lack, at the hadronic scale, of  (polarized) sea contributions 
which can play even a major r\^ole for $x \lesssim 0.3$ \cite{SVT97ss}.

All these limitations, however, do not obscure the main aim of our 
study, namely the discussion on the effects due the relativistic
requirement of the quark wave function.

Fig.~6 is devoted to the gluon distributions. Within scenario A the 
polarized gluon distributions remain {\it positive} at the experimental 
scale as a result  of the NLO evolution.
On the contrary a large amount of {\it negative} gluon polarization
is predicted within scenario B if one assumes anti-aligned gluons at the 
hadronic scale. The distribution is largely dependent on the  polarization 
fraction. The dotted line of Fig.~6a shows the consequence, in the deep 
inelastic regime, of the Jaffe's calculation at low energy. The absence of 
non-perturbative sea polarization results in a huge (and probably unrealistic) 
amount of negative gluon polarization to reproduce the neutron data. 
Such a large amount of gluons is, however, inconsistent with the
unpolarized gluon distributions as it is shown in Fig.~6b (dotted line again).

Fig.~7 shows the results of the analysis on the factorization scheme 
dependence. We have chosen to show both the proton and the more 
appreciable case of the polarized neutron data. The small difference 
gives us confidence in the results.

\subsubsection{Unpolarized parton distributions}

Figs.~8 and 9 complete our study of unpolarized partons (gluons are shown
in Fig.~6 as already mentioned). Particularly evident is the improvement,
due to the high momentum components introduced by the relativized solution,
in describing the proton $F_2$ distribution in the region $0.4 \lesssim x 
\lesssim 0.8$. The non-relativistic description is largely suppressed for 
$x > 0.4$ due to the lack of high momentum contributions. The solution of 
the mass 
operator (\ref{massop}) have the appropriate components to contribute in
this $x$-region . 
The work is completed by the comparison
of LO NLO evolution also in the case of unpolarized 
scattering (Fig.~8a) and the 
results on valence and sea distributions (Fig.~9). 
In this last case one could repeat 
the comments already formulated for the polarized 
distributions: the lack of non-perturbative sea
at the hadronic scale largely limits the possibility 
of reproducing the parton
distributions in the singlet sector. 

\section{Further Developments and conclusions}
\label{FDconcl}

The present paper is devoted to a semi-quantitative analysis of the effects 
due to relativistic covariance requirement in the study of polarized structure 
functions within constituent quark models. The investigation clearly shows the
relevance of relativistic effects due, 
in particular to the dynamics of the spin
observables in special relativity. 
In fact two new ingredients play a major r\^ole
in our light-front calculation: 
i) the presence of high momentum components and 
ii) the spin dynamics induced by the Melosh transformations.

\noindent i) The large amount of high 
momentum components in the nucleon wave function 
is generated by the solution of the relativized 
mass equation (\ref{massop}) and are 
relevant in reproducing the large $x$ 
behavior of polarized and unpolarized 
structure functions. The lack of such high 
momentum contributions have been 
repeatedly stressed  in connection with 
non-relativistic calculations \cite{noi1,highm} 
without reaching a satisfying solution within 
non-relativistic quark models.

\noindent ii) The Melosh rotation dynamics introduce the basic new  
ingredient in the calculations and its effect is quite sizeable in suppressing 
the proton response in the region $x \lesssim 0.4$.

The first natural extension of the present 
discussion is the study of transversity
distributions as they will be investigated 
by Drell-Yan processes. In fact relativity must
play a quite sizeable r\^ole for that 
observable \cite{jaffeji,sv98,cft99} and the 
light-front dynamics is a privileged tool to study such effects.

In addition to the advantages of our approach we want to stress also its 
limitations and approximations because a further study should remove them 
in order to provide a more quantitative study of the nucleon spin within the 
light-front dynamics. 

The most relevant limitation of the present investigation is probably the lack
of (non-perturbative) sea contributions at the hadronic scale due to the 
meson cloud of the nucleon. The effects can be rather large and significant
in particular for the spin response of the  
neutron. Work in this direction is
in progress and we want to extend 
the investigation of non-relativistic approaches
\cite{MairTr} inserting the recent successful description of the meson cloud
\cite{BorosThomas} within a coherent 
light-front formulation \cite{cftvprogress}

A second limitation is determined by the particular choice of the quark-quark 
interaction of Eq.~(\ref{massop}). As mentioned in few points of the present 
paper the large $x$ behavior of the nucleon response can be affected by the 
specific choice of the interquark interaction model. Work is in progress
to investigate the numerical solutions of more sophisticated hyper-central 
potentials already discussed in the non-relativistic limit \cite{TBM95}.
In addition we did not include  hyperfine interaction terms in our
analysis and we would like to extend the approach also to the consistent
light-front treatment of such interaction.

Finally the factorization scheme dependence is to be investigated 
in more detail
in connection with radiative model 
studies of the kind discussed in the present paper.


\section*{Acknowledgments}
We acknowledge useful conversations with S. Scopetta regarding factorization 
schemes and with F. Cano on the generalization of Eq.~(\ref{DVal}) to the
light-front approach. This work has been supported in part by
DGICYT-PB94-0080
and the TMR programme of the European Commission ERB FMRX-CT96-008. One of
us (VV) benefitted from a
travel grant from the Conselleria de Cultura, Educaci\'o i Ci\`encia de la
Generalitat Valenciana, which allowed his visit to DAPNIA-CEA, where this
work was partly done. He thanks J.P. Guichon and J.M. Laget for 
their efforts to make this visit possible and to all the members of the
Service for their hospitality.


\newpage

\protect
\begin{figure}

\begin{center}
\mbox{\epsfig{file=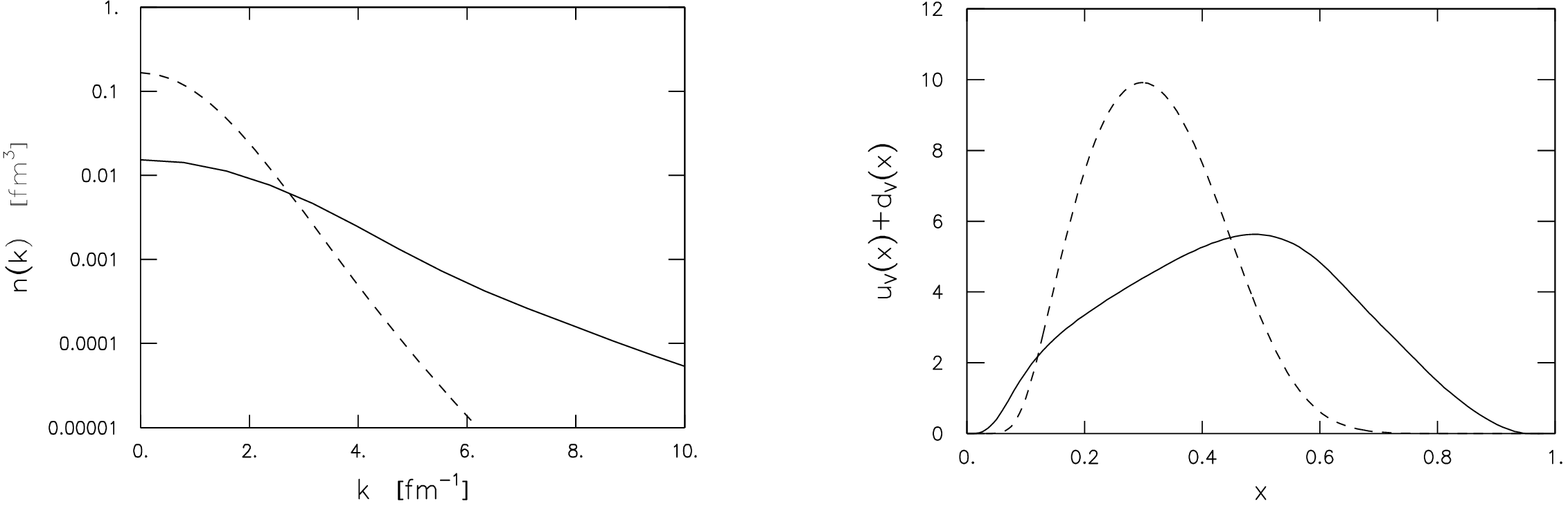,
width=0.35\linewidth,height=0.70\textheight, angle=-90}}
\end{center}
\vspace{3mm}
\caption{Fig.~1a (left panel): the valence momentum distribution of 
Eq.~(\ref{n_of_k}) as function of $|{\bf k}|$. The results of a full covariant
light-front calculation (full curve) are compared with the non-relativistic 
approximation (dashed curve).
The corresponding total valence distributions at the hadronic scale are shown on 
the right panel (Fig.~1b).}
\vspace{20mm}
\protect
\begin{center}
\mbox{\epsfig{file=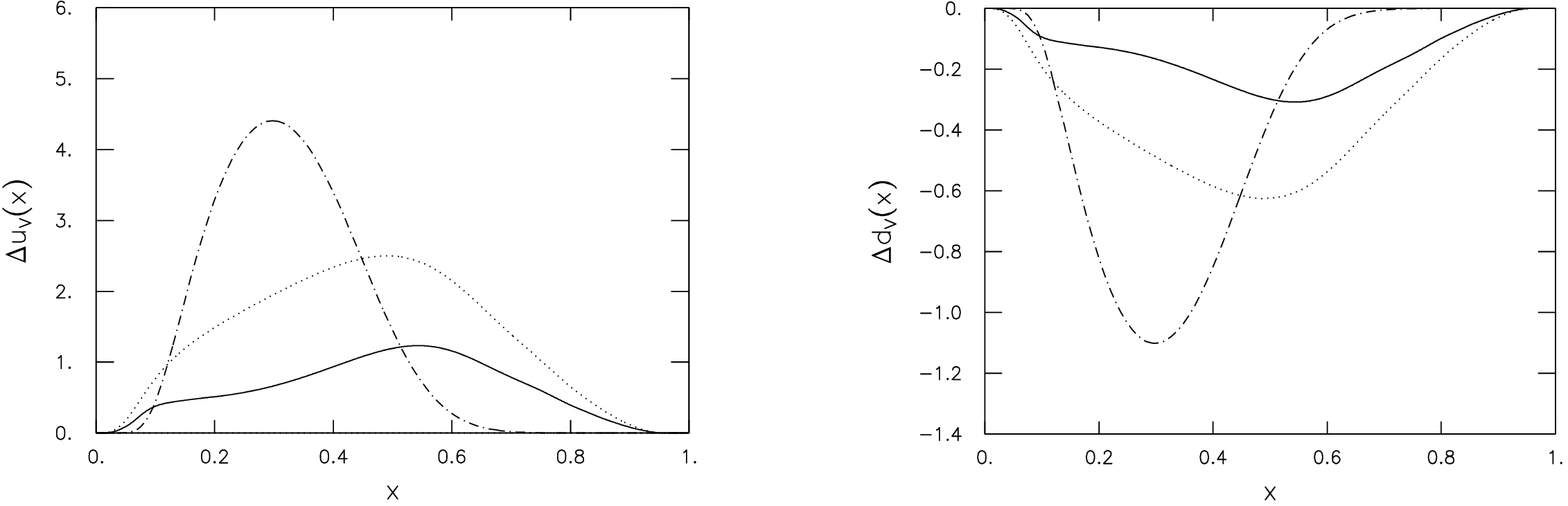,
width=0.35\linewidth,height=0.70\textheight, angle=-90}}
\end{center}
\vspace{3mm}
\caption{Left panel (Fig.~2a): the polarized distribution $\Delta
u_V(x,\mu_0^2)$
as function of $x$: the non-relativistic approximation (dot-dashed curve), and
the relativized solution of Eq.(\ref{massop})
which neglects Melosh rotations (dotted curve) are compared with the 
results of a complete light-front calculation (full curve). 
On the right panel the distibution $\Delta d_V(x,\mu_0^2)$ (same notation).}

\newpage

\protect
\begin{center}
\mbox{\epsfig{file=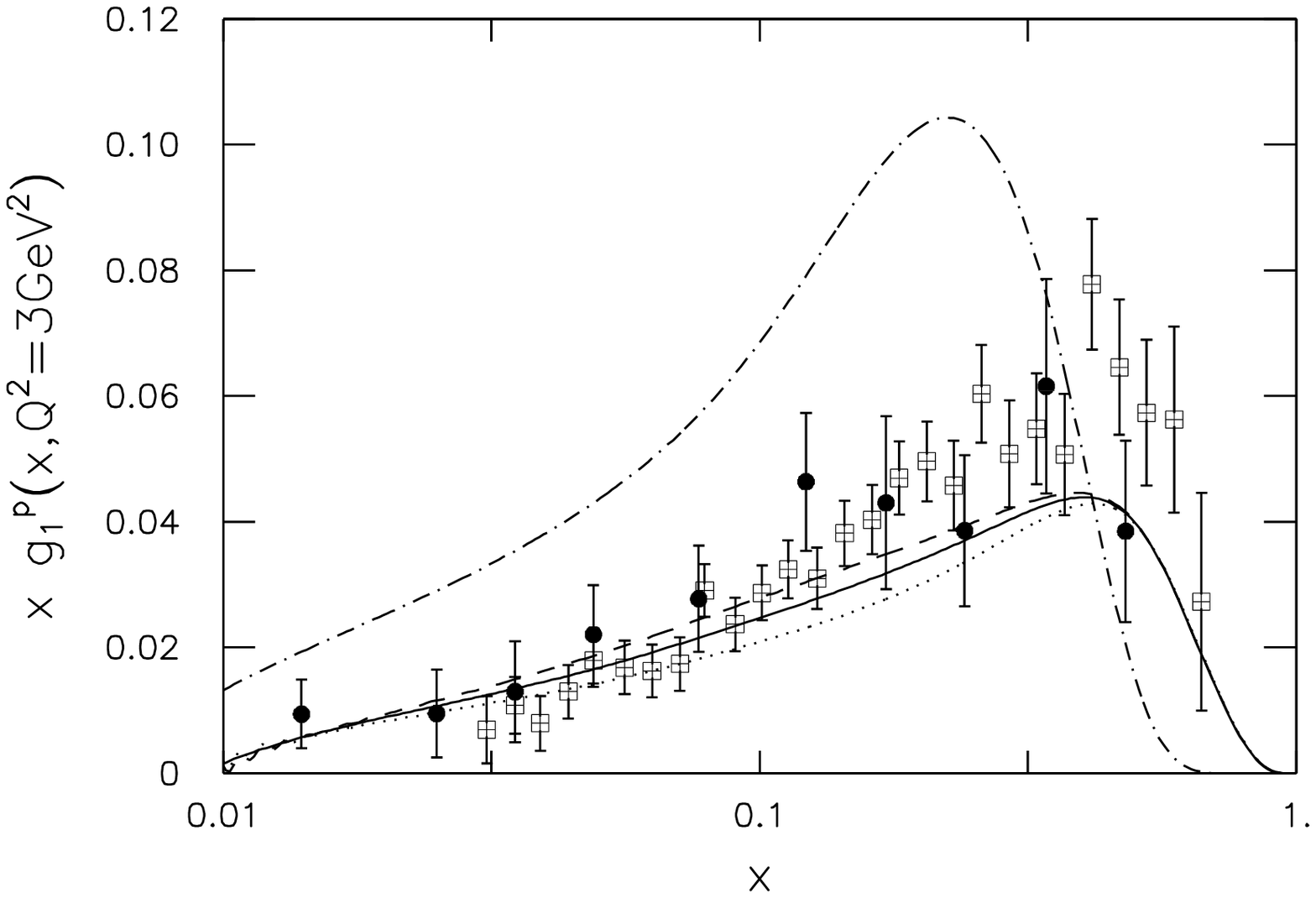,
width=0.70\linewidth,height=0.35\textheight, angle=0}}
\end{center}
\vspace{3mm}
\caption{The proton polarized structure function at $Q^2=3$ GeV$^2$.
The full curve represents the NLO ($\overline{MS}$) results of a complete light-front 
calculation within a scenario where no gluons are present at the hadronic scale (scenario A);
the corresponding non-relativistic calculation are shown by the dot-dashed line.
Scenario B is summarized by the dotted line in the case of negative polarized gluon
fraction ($\int \Delta G = -0.7$ as discussed in the text), and by the dashed
line in the case of positive gluon polarization ($\int \Delta G = +0.7$).
Data are from SMC [36], and SLAC(E142) [37] experiments.}
\vspace{20mm}
\protect
\begin{center}
\mbox{\epsfig{file=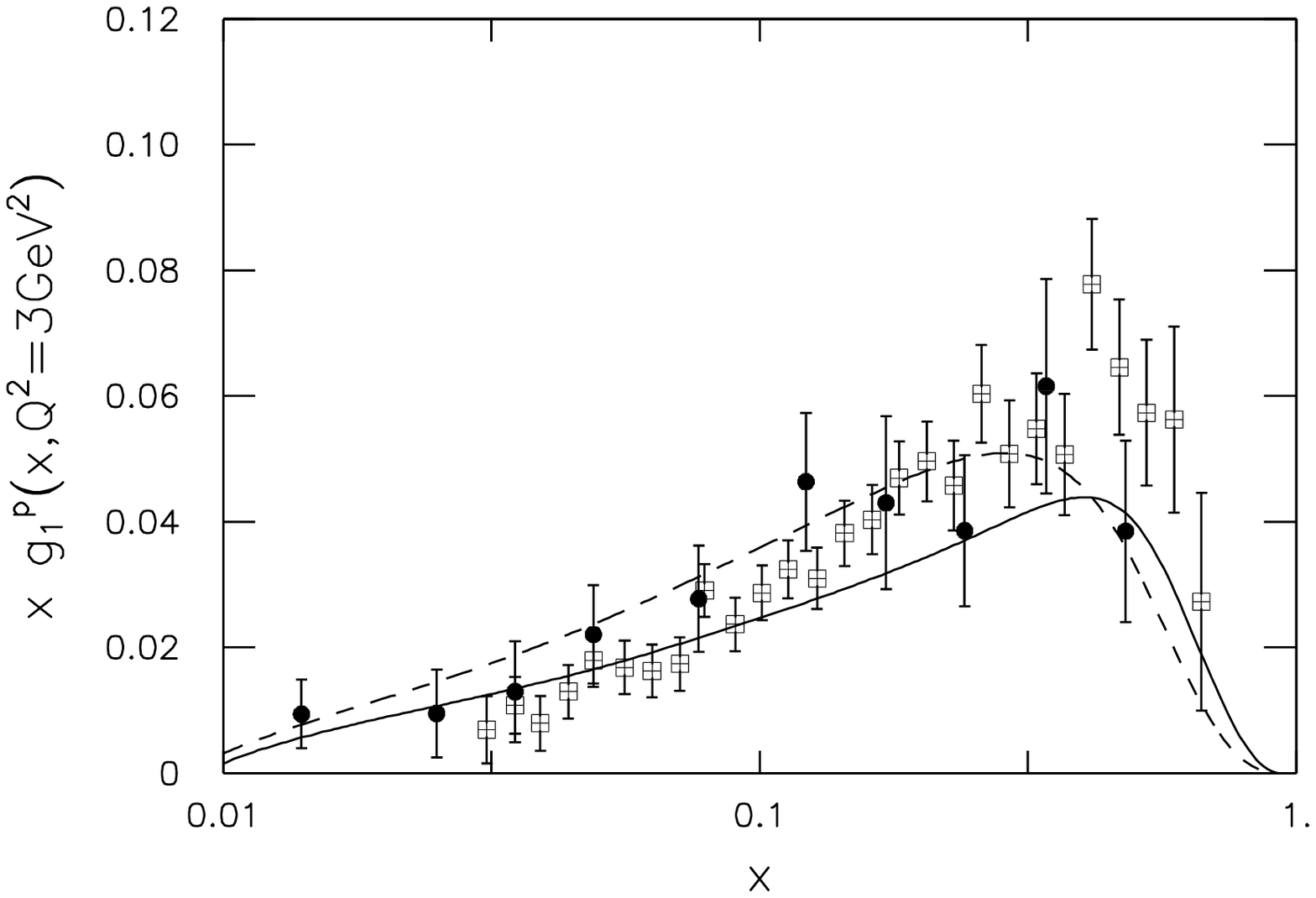,
width=0.70\linewidth,height=0.35\textheight, angle=0}}
\end{center}
\vspace{3mm}
\caption{The proton polarized structure function at $Q^2 = 3$ GeV$^2$
within scenario A. LO evolution is shown by the dashed curve, the NLO ($\overline{MS}$) 
results by the full curve. Data as in Fig.~3}

\newpage

\protect
\begin{center}
\mbox{\epsfig{file=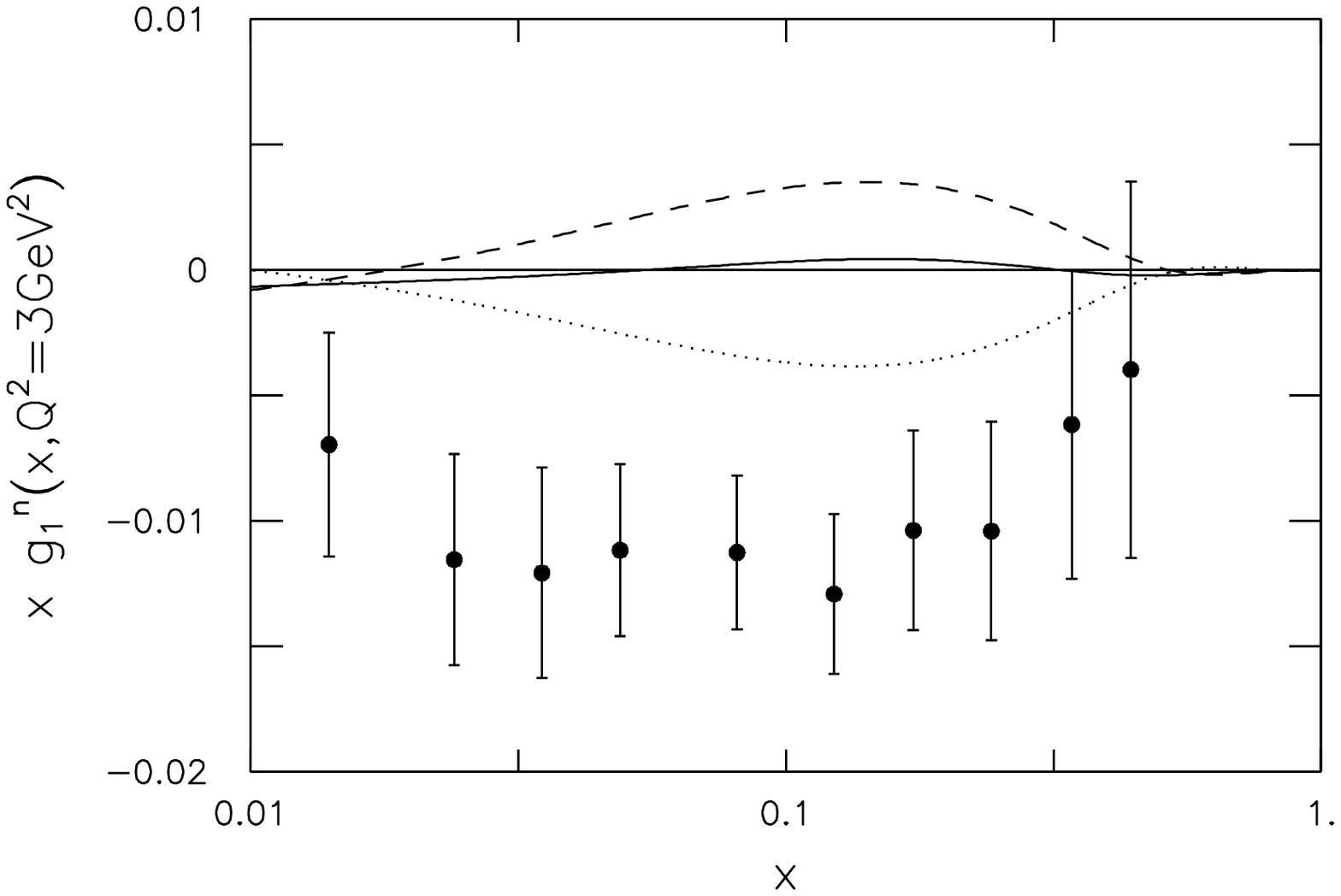,
width=0.40\linewidth,height=0.20\textheight, angle=0}}
\end{center}
\caption{The NLO ($\overline{MS}$) neutron polarized structure function at 
$Q^2=3$ GeV$^2$ within a complete light-front calculation.
The full curve represents the results within a scenario where no gluons are present 
at the hadronic scale (scenario A). Scenario B is summarized by the dotted line in 
the case of negative polarized gluon fraction ($\int \Delta G(x,\mu_0^2) = -0.7$ 
as discussed in the text), and by the dashed line in the case of positive gluon 
polarization ($\int \Delta G(x,\mu_0^2) = +0.7$). Data are from SLAC(E154) [38]
experiments.}
\protect
\begin{center}
\mbox{\epsfig{file=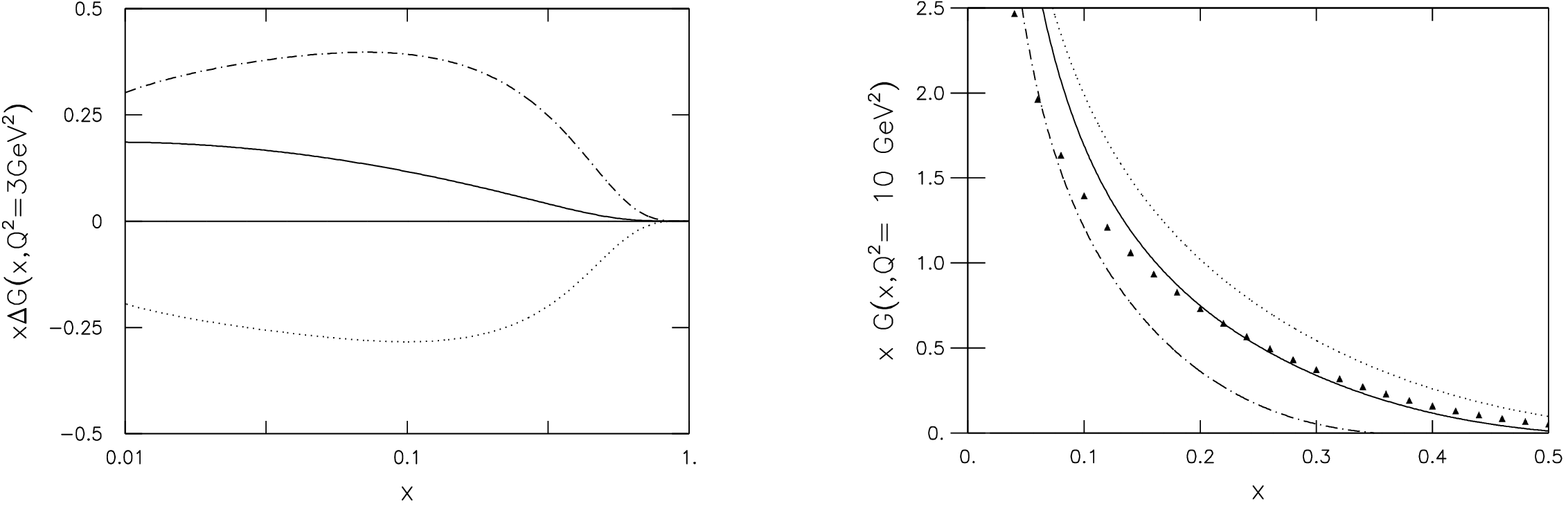,
width=0.35\linewidth,height=0.70\textheight, angle=-90}}
\end{center}
\caption{Left panel: Polarized gluon distributions at $Q^2=3$ GeV$^2$ obtained evolving 
at NLO ($\overline{MS}$) the polarized partons in both scenarios. Scenario A (full line).
Scenario B is summarized by the dotted line in the case of negative polarized gluon
fraction ($\int \Delta G(x,\mu_0^2) = -0.7$ as discussed in the text), and by the dot-dashed
line in the case of positive gluon polarization ($\int \Delta G(x,\mu_0^2) = +0.7$).\\
Right panel: unpolarized gluon distribution at $Q^2=10$ GeV$^2$ obtained evolving 
at NLO (DIS) unpolarized partons. Scenario A (full line); scenario B (dotted line). 
For comparison also the results at LO are shown in this case (dot-dashed).
CETQ4 NLO (DIS) fit of ref.[39]: full triangles.}

\newpage
\protect
\begin{center}
\mbox{\epsfig{file=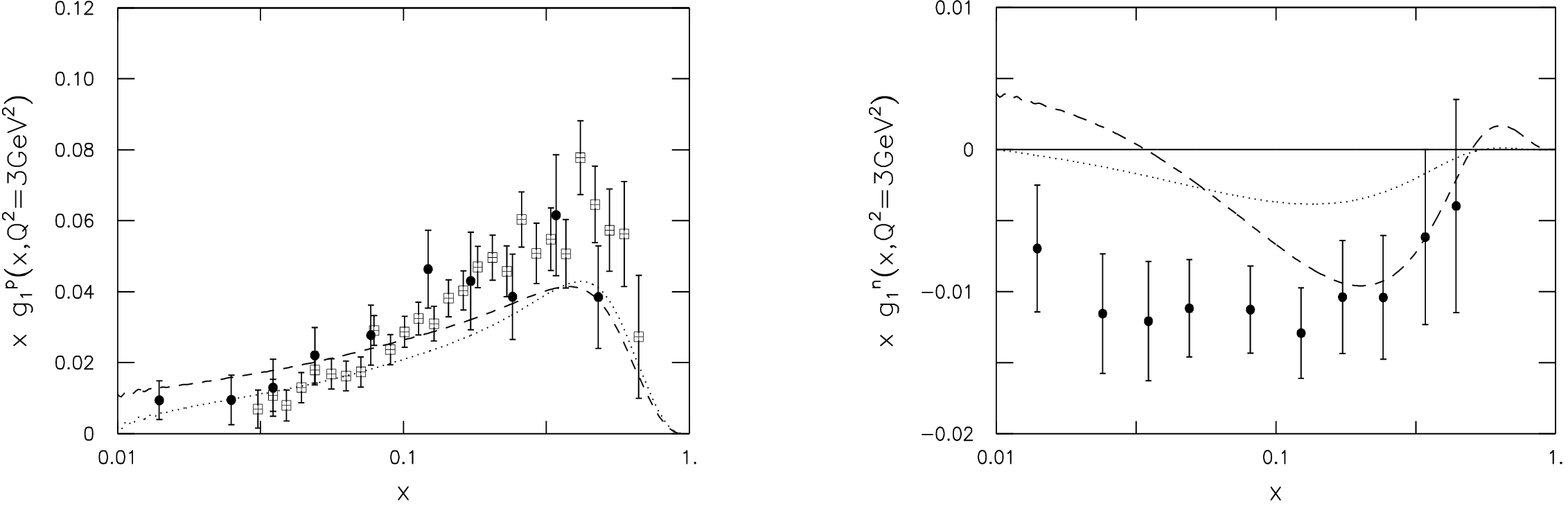,
width=0.35\linewidth,height=0.70\textheight, angle=-90}}
\end{center}
\vspace{3mm}
\caption{Scheme dependence of the polarized proton and neutron structure functions 
at $Q^2=3$ GeV$^2$. $\overline {MS}$ dotted lines; $AB$: dashed lines.}
\vspace{20mm}
\protect
\begin{center}
\mbox{\epsfig{file=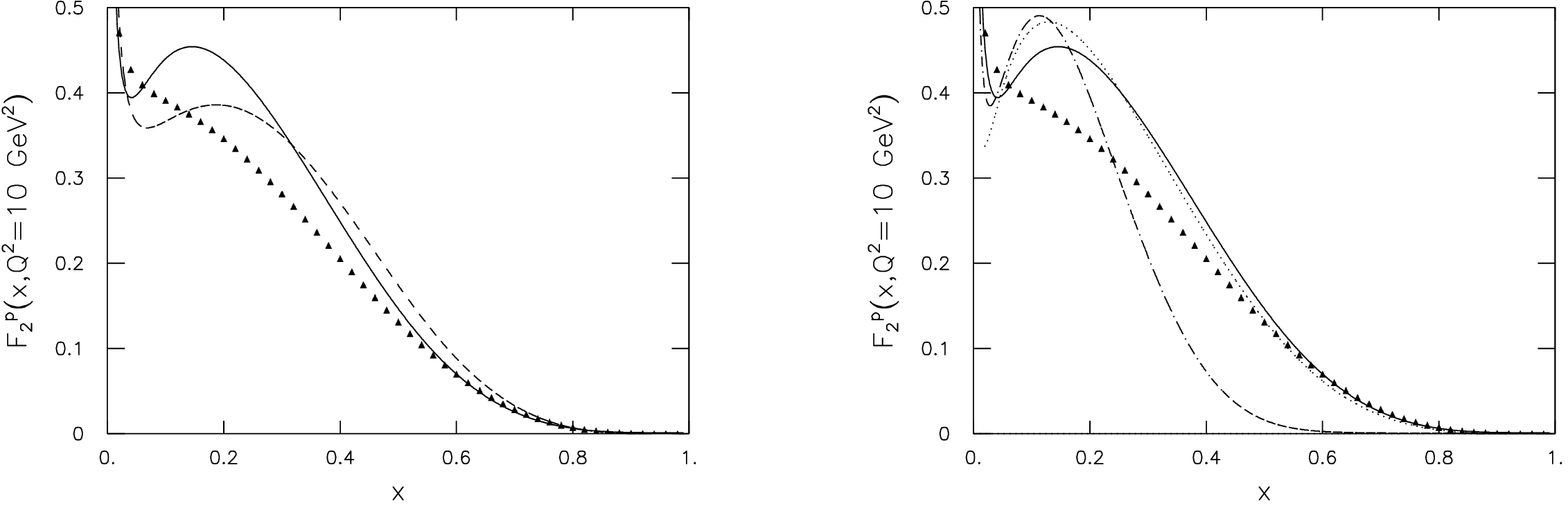,
width=0.35\linewidth,height=0.70\textheight, angle=-90}}
\end{center}
\vspace{3mm}
\caption{Left panel: unpolarized $F_2$ structure function at $Q^2=10$ GeV$^2$ as 
function of $x$. LO evolution (dashed) NLO (DIS) full line. 
Right panel: the $F_2$ structure function of the proton at $Q^2=10$ GeV$^2$ as
predicted by the full light-front calculation at NLO (full curve: scenario A, 
dotted curve: scenario B) is compared with the non-relativistic approximation 
(dot-dashed curve).
CETQ4 fit of ref.[39]: full triangles.}
\vspace{20mm}
\protect
\begin{center}
\mbox{\epsfig{file=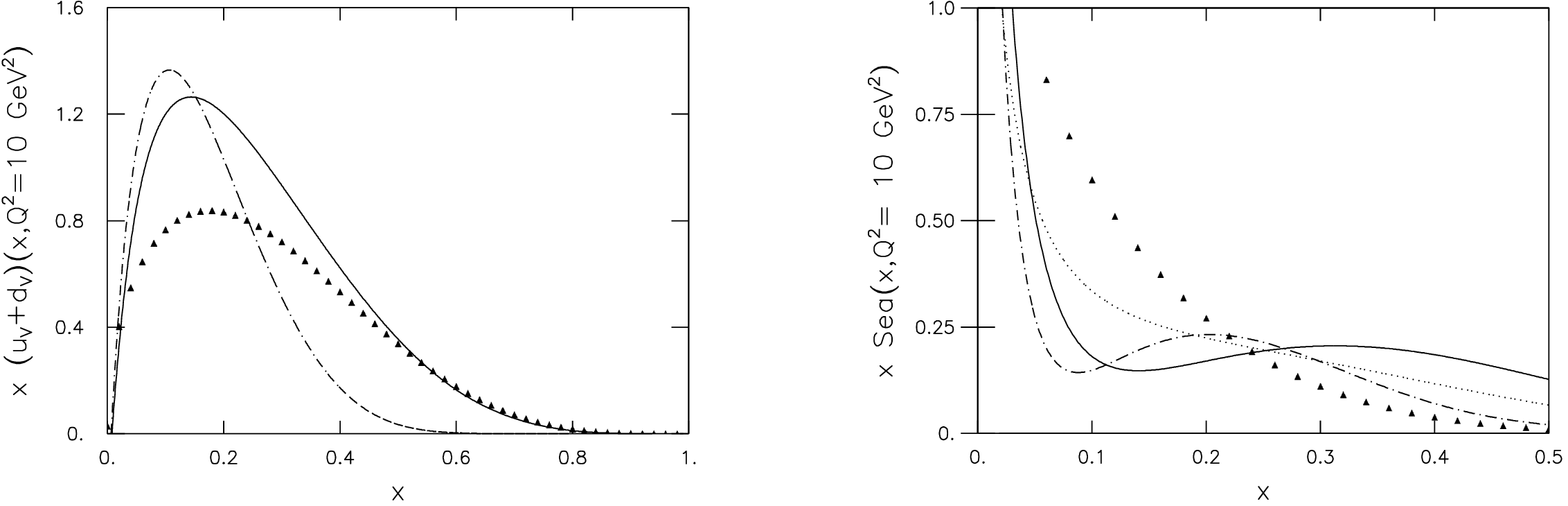,
width=0.35\linewidth,height=0.70\textheight, angle=-90}}
\end{center}
\vspace{3mm}
\caption{Left panel: unpolarized NLO (DIS) valence parton distribution at
$Q^2=10$ GeV$^2$ as fuction of $x$. Light-front calculations: 
full curve; non-relativistic approximation (dot-dashed curve); 
NLO (DIS) CTEQ4 fit of ref.[39]: full triangles.
Right panel: total unpolarized NLO (DIS) sea distribution at at $Q^2=10$ GeV$^2$.
Light-front results: scenario A (full curve), scenario B (dotted curve), 
non-relativistic approximation (dot-dashed curve). }

\end{figure}

\end{document}